# Moment based Spectrum Sensing Algorithm for Cognitive Radio Networks with Noise Variance Uncertainty

Tadilo Endeshaw Bogale and Luc Vandendorpe
ICTEAM Institute
Universitè catholique de Louvain
Place du Levant, 2, B-1348, Louvain La Neuve, Belgium
Email: {tadilo.bogale, luc.vandendorpe}@uclouvain.be

*Abstract*— This paper proposes simple moment based spectrum sensing algorithm for cognitive radio networks in a flat fading channel. It is assumed that the transmitted signal samples are binary (quadrature) phase-shift keying BPSK (QPSK), M-ary quadrature amplitude modulation (QAM) or continuous uniformly distributed random variables and the noise samples are independent and identically distributed circularly symmetric complex Gaussian random variables all with unknown (imperfect) variance. Under these assumptions, we propose a simple test statistics employing a ratio of fourth and second moments. For this statistics, we provide analytical expressions for both probability of false alarm ($P_f$) and probability of detection ($P_d$) in an additive white Gaussian noise (AWGN) channel. We confirm the theoretical expressions by computer simulation. Furthermore, under noise variance uncertainty, simulation results demonstrate that the proposed moment based detector gives better detection performance compared to that of energy detector in AWGN and Rayleigh fading channels.

## I. Introduction

The current wireless communication networks adapt fixed spectrum access strategy. The Federal Communications Commission have found that the fixed spectrum access strategy utilizes the available frequency bands inefficiently [1], [2]. A promising approach of addressing this problem is to deploy a cognitive radio (CR) network. One of the key characteristics of a CR network is its ability to discern the nature of the surrounding radio environment. This is performed by the spectrum sensing (signal detection) part of a CR network.

The most common spectrum sensing algorithms for CR networks are matched filter, energy and cyclostationary based algorithms. If the characteristics of the primary user such as modulation scheme, pulse shaping filter and packet format are known perfectly, matched filter is the optimal signal detection method as it maximizes the received Signal-to-Noise Ratio (SNR). In practice, this information can be known a priori. The main drawback of matched filter detector is that it needs dedicated receiver to detect each signal characteristics of a primary user [3]. Energy detector does not need any information about the primary user and it is simple to implement. However, energy detector is very sensitive to noise variance uncertainty, and there is an SNR wall below which energy detector is not able to guarantee a certain detection performance [3]–[5]. Cyclostationary based detection method is robust against noise variance uncertainty and it can reject the effect of adjacent channel interference. However, the computational complexity of this detection method is very high, and large number of samples are required to exploit the cyclostationarity nature of the received samples [5], [6]. On the other hand, this method is not robust against cyclic frequency offset which can occur due to clock mismatch between the transmitter and receiver [7]. In [8], Eigenvalue-based spectrum sensing algorithm has been proposed. This algorithm is robust against noise variance uncertainty. However, the computational complexity of this method is very high.

In a conventional digital communication system, the transmitted signal samples are taken from a given constellation. This constellation may be binary (quadrature) phase-shift keying BPSK (QPSK) or M-ary quadrature amplitude modulation (QAM). In all constellations, each component (either real or imaginary) of a sample takes a value in between $[-b, b], b > 0$, where $b$ depends on the SNR of the received signal. For these reasons, we assume that the transmitted signal samples are BPSK, QPSK, M-ary QAM or continuous uniformly distributed random variables[1]. We also assume that the noise samples are independent and identically distributed (i.i.d) circularly symmetric complex Gaussian (ZMCSCG) random variables all with unknown (imperfect) variance. Under these assumptions, we show that the ratio of the fourth absolute moment and the square of second absolute moment give 2 and $< 2$, in noise only and signal plus noise cases, respectively. Due to this, we propose a test statistics as 2 minus the ratio of the fourth absolute moment and the square of second absolute moment. For this test statistics, we provide analytical expressions for both probability of false alarm ($P_f$) and probability of detection ($P_d$) in an additive white Gaussian noise (AWGN) channel environment. As the $P_f$ expression

The authors would like to thank SES for the financial support of this work, the french community of Belgium for funding the ARC SCOOP and BELSPO for funding the IAP BESTCOM project.

---

[1]We would like to mention here that in the case of Orthogonal Frequency Division Multiplexing (OFDM) signals, the scenario mentioned in this paragraph can be exploited by considering the fourier transform of the received samples.

of the proposed detector is not dependent on the actual noise variance, the proposed detector is robust against noise variance uncertainty. We also confirm the theoretical expressions by computer simulations. Furthermore, we demonstrate by computer simulations that the proposed moment based detector gives better detection performance compared to that of the well known energy detector in AWGN and Rayleigh fading channels.

This paper is organized as follows: Section II discusses the hypothesis test problem. In Section III, some preliminary results on moments for random variables are discussed. Section IV presents the proposed moment based detector. In Section V, computer simulations are used to compare the performance of the proposed moment based detector to that of energy detector. Conclusions are presented in Section VI.

## II. PROBLEM FORMULATION

Let $\mathbf{s} = \{s[n]\}_{n=1}^{N}$ denote the transmitted discrete time baseband signal vector. If we assume an AWGN channel, the observed baseband signal has the following form [9]

$$y[n] = \begin{cases} s[n] + w[n], & H_1 \\ w[n], & H_0 \end{cases} \quad n = 1, \cdots, N \quad (1)$$

where $s[n]$, $w[n]$ and $N$ are the $n$th transmitted signal sample, $n$th noise sample and number of samples, respectively. The noise samples $\{w[n]\}_{n=1}^{N}$ are assumed to be i.i.d ZMCSCG random variables[2]. The variance of each component of $w[n]$ is assumed to be $\sigma^2$ which is unknown or known imperfectly. The aim of a CR spectrum sensing is to detect the presence or absence of the transmitted signal samples $\{s[n]\}_{n=1}^{N}$.

## III. PRELIMINARY

The $k$th moment of a random variable $X$ is defined as [10]

$$M_k = \mathrm{E}\{X^k\} = \begin{cases} \sum_x x^k p(x), & \text{For discrete X} \\ \int_{-\infty}^{\infty} x^k p(x) dx, & \text{For continuous X} \end{cases} \quad (2)$$

where $\mathrm{E}\{.\}$ and $p(.)$ denote expectation and probability density function, respectively.

For a discrete uniform random variable $X$ with $P$ possible values in $[-b, \ b]$, the $k$th moment is thus given by

$$M_k = \frac{b^k(-1)^k}{P(P-1)^k} \sum_{i=0}^{P-1}(P - 2i - 1)^k. \quad (3)$$

For a continuous uniform random variable $X \sim \mathcal{U}[-b, \ b]$, applying (2) gives

$$M_k = \begin{cases} \frac{b^k}{k+1}, & \text{For even k} \\ 0, & \text{For odd k.} \end{cases} \quad (4)$$

For a continuous Gaussian random variable $X \sim \mathcal{N}(0, \sigma^2)$, applying (2) yields

$$M_k = \begin{cases} 1 \times 3 \times \cdots \times (k-1)\sigma^k, & \text{For even k} \\ 0, & \text{For odd k.} \end{cases} \quad (5)$$

[2]In the case of nonzero mean received signal samples $\{y[n]\}_{n=1}^{N}$, one can remove the mean from the received samples.

## IV. ABSOLUTE MOMENT BASED DETECTOR

The $k$th absolute moment of a random variable $X$ is defined as $M_{axk} \triangleq \mathrm{E}\{|X|^k\}$. By employing (3) - (5), one can get the following second and fourth absolute moments

$$\begin{aligned}
M_{ay2} &= \mathrm{E}\{|y[n]|^2\} \\
&= 2\sigma^2(\beta + 1), \qquad \text{Any s[n]} \\
M_{ay4} &= \mathrm{E}\{|y[n]|^4\} \\
&= \sigma^4(4\beta^2 + 16\beta + 8), \qquad \text{s[n] = BPSK, QPSK} \\
&= \sigma^4(\frac{132}{25}\beta^2 + 16\beta + 8), \quad \text{s[n] = 16 QAM } (P = 4) \\
&= \sigma^4(\frac{116}{21}\beta^2 + 16\beta + 8), \quad \text{s[n] = 64 QAM } (P = 8) \\
&= \sigma^4(\frac{28}{5}\beta^2 + 16\beta + 8), \quad \text{s[n] = CU}
\end{aligned} \quad (6)$$

where $\beta = \frac{\mathrm{E}\{|s[n]|^2\}}{\mathrm{E}\{|w[n]|^2\}}$ is the SNR of the received signal samples and CU stands for continuous uniform random variable. As can be seen from this equation, the fourth moment gap between an M-ary and a CU random variable signal decreases as M increases. Thus, without loss of generality, one can apply the results of a continuous uniform random variable for higher modulation orders (for example 512 QAM signals). Hence, the above expressions can represent the second and fourth moments of practically relevant signal constellations. The ratio $T \triangleq -\frac{M_{ay4}}{M_{ay2}^2}$ is computed as

$$\begin{aligned}
T &= -\frac{M_{ay4}}{M_{ay2}^2} \qquad\qquad\qquad\qquad (7) \\
&= -2, \qquad\qquad\qquad \text{Noise only} \\
&= -2 + \left(\frac{\beta}{\beta+1}\right)^2, \qquad \text{s[n] = BPSK, QPSK} \\
&= -2 + \frac{17}{25}\left(\frac{\beta}{\beta+1}\right)^2, \qquad \text{s[n] = 16 QAM} \\
&= -2 + \frac{13}{21}\left(\frac{\beta}{\beta+1}\right)^2, \qquad \text{s[n] = 64 QAM} \\
&= -2 + \frac{3}{5}\left(\frac{\beta}{\beta+1}\right)^2, \qquad \text{s[n] = CU.}
\end{aligned}$$

However, since $M_{ay2}$ and $M_{ay4}$ are not known a priori, we employ their estimated values which can be computed as

$$\widehat{M_{ayk}} = \frac{1}{N}\sum_{n=1}^{N}|y[n]|^k, \qquad k = 2, 4. \quad (8)$$

And the estimated $T$ becomes

$$\widehat{T} = -\frac{\widehat{M_{ay4}}}{\widehat{M_{ay2}^2}}. \quad (9)$$

Thus, the binary hypothesis test of (1) turns to examining whether $\widehat{T} = -2$ or $\widehat{T} > -2$. To get the exact test statistics, $P_d$ and $P_f$ expressions, we examine the following Theorem [11].

*Theorem 1*: Given a real valued function $\widehat{T} = -\frac{\widehat{M_{ay4}}}{\widehat{M_{ay2}^2}}$, the asymptotic distribution of $\sqrt{N}(\widehat{T} - T)$ is

$$\sqrt{N}(\widehat{T} - T) \sim \mathcal{N}(0, \tilde{\sigma}^2) \quad (10)$$

where $\tilde{\sigma}^2 = \mathbf{v}\mathbf{\Phi}\mathbf{v}^T$,

$$\mathbf{v} = \left[\frac{\partial \widehat{T}}{\partial \widehat{M}_{ay2}}, \frac{\partial \widehat{T}}{\partial \widehat{M}_{ay4}}\right]_{\widehat{M}_{ay2}=M_{ay2}, \widehat{M}_{a4}=M_{ay4}}$$
$$= \left[2\frac{M_{ay4}}{M_{ay2}^3}, -\frac{1}{M_{ay2}^2}\right] \quad (11)$$

and $\mathbf{\Phi}$ is the asymptotic covariance matrix of a multivariate random variable $\sqrt{N}([\widehat{M}_{a2}, \widehat{M}_{a4}]^T - [M_{a2}, M_{a4}]^T) \sim \mathcal{N}(\mathbf{0}, \mathbf{\Phi})$.

*Proof:* See *Theorem* 3. 3. A on page 122 of [11]. ∎

By applying multivariate central limit theorem [12], it can be shown that $\mathbf{\Phi}_{(1,1)} = M_{ay4} - M_{ay2}^2$, $\mathbf{\Phi}_{(1,2)} = \mathbf{\Phi}_{(2,1)} = M_{ay6} - M_{ay2}M_{ay4}$ and $\mathbf{\Phi}_{(2,2)} = M_{ay8} - M_{ay4}^2$. Substituting $\mathbf{\Phi}$ into (10) gives

$$\tilde{\sigma}^2 = \frac{4M_{ay4}^2 \mathbf{\Phi}_{(1,1)} - 4M_{ay4}M_{ay2}\mathbf{\Phi}_{(1,2)} + M_{ay2}^2 \mathbf{\Phi}_{(2,2)}}{M_{ay2}^6}$$
$$= \frac{4M_{ay4}^3 + M_{ay2}^2 M_{ay8} - 4M_{ay2}M_{ay4}M_{ay6} - M_{ay2}^2 M_{ay4}^2}{M_{ay2}^6}. \quad (12)$$

After several steps, one can get the following variances

$$\tilde{\sigma}^2 = 4, \quad (13)$$
$$= \frac{8\beta^4 + 32\beta^3 + 40\beta^2 + \kappa}{(\beta+1)^6},$$
$$= \frac{0.234\beta^6 + 2.765\beta^5 + 17.114\beta^4 + 42.24\beta^3 + 46.4\beta^2 + \kappa}{(\beta+1)^6},$$
$$= \frac{0.26\beta^6 + 3.51\beta^5 + 19.49\beta^4 + 44.8\beta^3 + 47.62\beta^2 + \kappa}{(\beta+1)^6},$$
$$= \frac{0.325\beta^6 + 3.977\beta^5 + 20.503\beta^4 + 45.715\beta^3 + 48\beta^2 + \kappa}{(\beta+1)^6},$$

where $\kappa = 24\beta + 4$ and the first, second, third, fourth and fifth equalities are for Noise only, BPSK (QPSK), 16 QAM, 64 QAM and CU random variable scenarios, respectively.

Thus, as $T = -2$ under $H_0$ hypothesis, we propose the following test statistics

$$T_s = \sqrt{N}(\widehat{T} + 2). \quad (14)$$

Using the test statistics (14), we decide $\{y[n]\}_{n=1}^N$ of (1) as $H_0$ if $T_s < \lambda$ and as $H_1$ if $T_s \geq \lambda$, where $\lambda$ is a threshold value that is chosen to guarantee a certain performance. In general, $\lambda$ is selected such that (14) can guarantee either a constant $P_d$ or $P_f$.

Mathematically, $P_f(\lambda)$ is expressed as

$$P_f(\lambda) = Pr\{T_s > \lambda | H_0\}. \quad (15)$$

Under $H_0$ hypothesis, as $T_s \sim \mathcal{N}(0, 4)$, $P_f$ becomes

$$P_f = \int_\lambda^\infty \frac{1}{\sqrt{2\pi\tilde{\sigma}^2}} \exp^{-\frac{x^2}{2\tilde{\sigma}^2}} dx = Q\left(\frac{\lambda}{\sqrt{\tilde{\sigma}^2}}\right) = Q\left(\frac{\lambda}{2}\right) \quad (16)$$

where $Q(.)$ is the Q-function which is defined as [13]

$$Q(\lambda) = \frac{1}{\sqrt{2\pi}} \int_\lambda^\infty \exp^{-\frac{x^2}{2}} dx.$$

Mathematically, $P_d(\lambda)$ is expressed as

$$P_d(\lambda) = Pr\{T_s > \lambda | H_1\} \quad (17)$$

Under $H_1$ hypothesis, $T_s \sim \mathcal{N}(\mu, \tilde{\sigma}^2)$ where $\mu = \sqrt{N}(T+2)$. As a result[3]

$$P_d(\lambda) = \int_\lambda^\infty \frac{1}{\sqrt{2\pi\tilde{\sigma}^2}} \exp^{-\frac{(x-\mu)^2}{2\tilde{\sigma}^2}} dx = Q\left(\frac{\lambda - \mu}{\sqrt{\tilde{\sigma}^2}}\right). \quad (18)$$

From this expression, one can understand that for a given $\beta > 0$ (i.e., SNR) and $\lambda$, increasing $N$ increases $P_d$. This is due to the fact that $Q(.)$ is a decreasing function. Thus, the proposed moment based detector is consistent. Moreover, as $\mu$ and $\sigma$ are not the same in all modulation schemes, the $P_d$ of different modulation schemes may not be equal.

We would like to mention here that similar approach has been proposed in [14] to detect constant amplitude signals. Therefore, the current work can be considered as a generalized version of [14]. However, as such generalization needs derivations of several parameters to get the variances (13) for different modulation schemes, our work is not a straightforward extension of [14].

## V. SIMULATION RESULTS

In this section, we provide simulation results for the proposed moment based and energy detectors. All results of this section are obtained by averaging 10000 experiments.

### A. Verification of theoretical expressions

In this subsection, we verify the theoretical $P_d$ and $P_f$ expressions of the moment based detector (14) for different modulation types under an AWGN channel environment. For this simulation, we assume that the noise variance is known perfectly[4]. As can be seen from Fig. 1, the theoretical $P_d$ and $P_f$ matches with that of the simulation for all modulation schemes. Furthermore, for fixed $P_f$, $P_d$ decreases as the modulation order increases. This is due to the fact that $\mu$ decreases as the modulation order increases and at the simulated SNR (i.e., -10dB), $\tilde{\sigma}^2$ of all modulation schemes are almost the same (see (13)).

### B. Comparison of moment and energy based detectors

In this subsection, we compare the performance of the energy and moment based detectors for pulse shaped transmitted signals under noise variance uncertainty. The comparison is performed in both AWGN and Rayleigh fading channels by setting $P_f = 0.1$ and $N = 2^{16}$. It is assumed that the transmitter and receiver employ a square root raised cosine filter with roll-off factor $0.2$. The over-sampling factor and filter length are set to $S = 4$ and $L = 4S + 1$, respectively. According to [4], in an uncertain noise variance signal detection algorithm, the actual noise variance can be modeled as a bounded interval of $[\frac{1}{\epsilon}\sigma^2 \; \epsilon\sigma^2]$ for some $\epsilon = 10^{\Delta\sigma^2/10} > 1$, where the uncertainty $\Delta\sigma^2$ is expressed in dB. We assume that this bound follows a uniform distribution, i.e., $\mathcal{U}[\frac{1}{\epsilon}\sigma^2 \ \epsilon\sigma^2]$. The SNR is defined as $SNR \triangleq \sigma_s^2/\sigma^2$, where $\sigma_s^2$ is the variance of $\{s[n]\}_{n=1}^N$. The noise variance is the same for one observation (since it has a short duration) and follow a uniform distribution during several observations. Moreover, in a Rayleigh fading channel, the channel is constant for one observation and follows a Rayleigh distribution during several observations. For better exposition, QPSK and 16 QAM signals are considered.

---
[3]As we can see from (7), one can get $\mu = 0$ and $\mu > 0$ under $H_0$ and $H_1$ hypothesis, respectively.
[4]Here perfect noise variance is required just to get $P_d$ (depends on SNR).

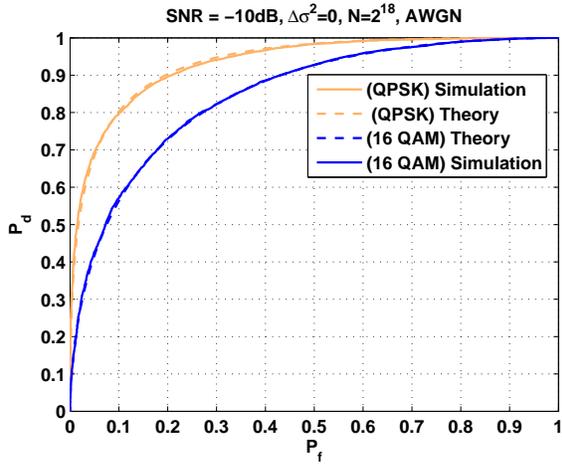

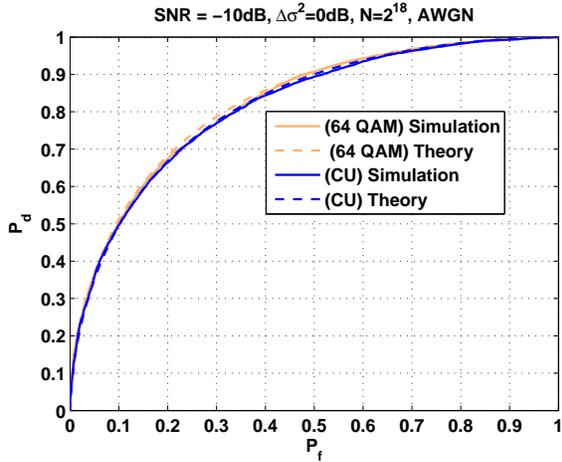

Fig. 1. Theoretical and simulated $P_d$ and $P_f$ of the proposed moment based detector in AWGN channel for (a) QPSK and 16 QAM, (b) 64 QAM and CU signals.

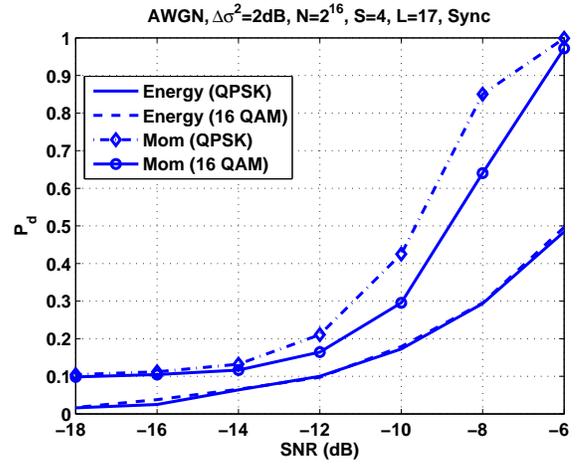

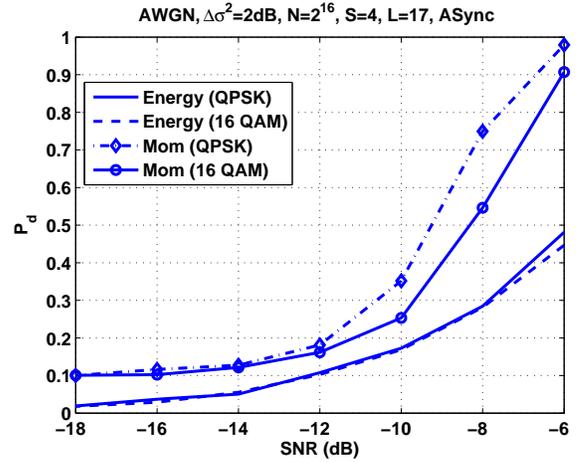

Fig. 2. Comparison of the proposed moment based (Mom) and energy (Energy) detectors under noise variance uncertainty in AWGN channel (a) Perfectly synchronized receiver, (b) Asynchronous receiver (i.e., with bit synchronization errors)

Fig. 2 and Fig. 3 show the performance of energy and moment based detectors under noise variance uncertainty with synchronous and asynchronous (i.e., with bit synchronization errors) receiver scenarios. From these figures, we can observe that the proposed moment based detector achieves better detection performance compared to that of energy detector for all scenarios. Moreover, the proposed detector achieves the best performance when the transmitter and receiver are synchronized perfectly. As expected, the performance of moment based detector decreases as the modulation order increases in both AWGN and Rayleigh fading channels. And the performance of energy detector is not affected by bit synchronization errors [15].

## VI. CONCLUSIONS

This paper proposes simple moment based spectrum sensing algorithm for cognitive radio networks in flat fading channels. We assume that the transmitted signal samples are BPSK, QPSK, M-ary QAM or continuous uniformly distributed random variables and the noise samples are independent and identically distributed circularly symmetric complex Gaussian random variables all with unknown (imperfect) variance. Under these assumptions, we propose a simple test statistics employing a ratio of fourth and second moments. For this test statistics, we provide analytical expressions for both $P_f$ and $P_d$ in an AWGN channel environment. Furthermore, under noise variance uncertainty, we demonstrate by computer simulation

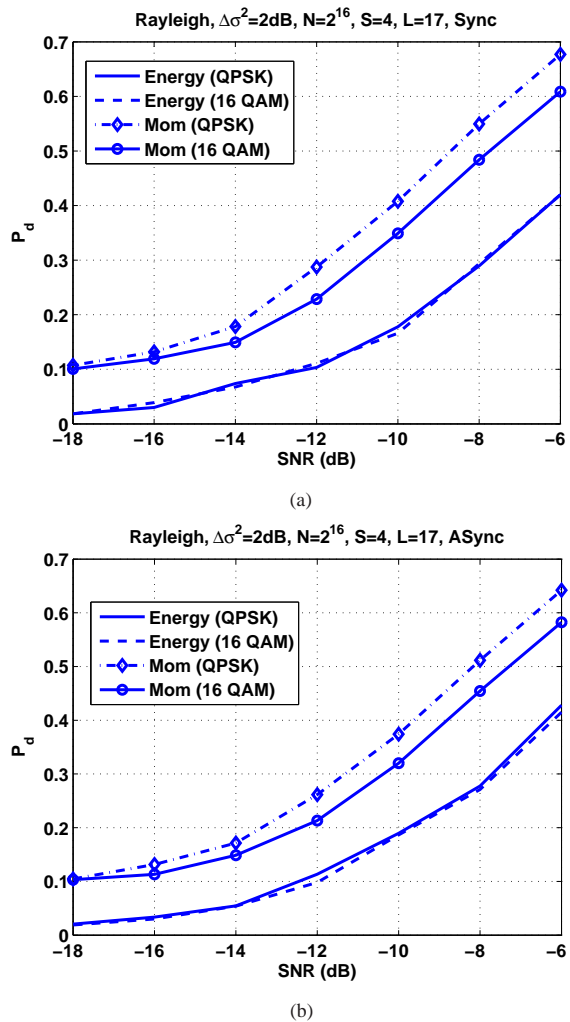

Fig. 3. Comparison of the proposed moment based (Mom) and energy (Energy) detectors under noise variance uncertainty in Rayleigh fading channel (a) Perfectly synchronized receiver, (b) Asynchronous receiver (i.e., with bit synchronization errors)

results that the proposed moment based detector gives better detection performance compared to that of the well known energy detector in AWGN and Rayleigh fading channels.